# Emission Transfer of Interstitial Atoms Under Shock Deformation of a Metal Surface

A. I. Karasevskii, A. Yu. Naumuk

*G. V. Kurdyumov Institute for Metal Physics, 36 Vernadsky Boulevard, Kyiv 03142, Ukraine*

The process of anomalous transfer of interstitial atoms during impact deformation of the crystal surface is described theoretically. As shown that surface impact leads to the formation of a wave of inhomogeneous atomic displacements in the medium, which propagates from the surface into the depth of the crystal. The formation of a deformation wave leads to a change in the interatomic distance at the wave front and a change in the potential energy for interstitial atoms. Interstitial atoms at the front of the deformation wave receive an additional impulse, which leads to an increase in their kinetic energy and contributes their movement deep into the crystal.

## I. INTRODUCTION

In a series of works Ref. 1–3, a significant increase in the mass transfer rate of interstitial atoms in a metal was found under a shock load on its surface. In this case, the impact action can be of a different nature - mechanical shocks (Ref. 1), laser radiation pulses (Ref. 2), spark discharges (Ref. 3), and the like.

A number of models have been proposed for the theoretical description of anomalous mass transfer: a mechanism of interstitial atom migration under high-rate strain influence was presented by Gurevich et al. (Ref. 1); it was assumed in Ref. 4–6 that an interstitial atom is captured by a dislocation and transported by it to macroscopic distances deep into the crystal; It was shown in Ref. 7 that, at the front of a deformation wave, a significant decrease in the activation barrier of transfer for interstitial atoms of a crystal can occur.

In this work, the anomalous transfer of interstitial atoms under a shock load on its surface is associated with the formation of a deformation wave of longitudinal atomic displacements in a crystal. It is shown that a surface impact leads to the formation of a wave of inhomogeneous atomic displacements in the medium, which propagates from the surface to the depth of the crystal.

In this case, due to a change in the interatomic distance at the front of the deformation wave of the crystal, the potential energy barrier for interstitial atoms decrees. In addition, interstitial atoms on the front of the deformation wave receive an additional impulse, which can lead to contributes to the activationless overcoming of interstitial potential barriers by the atoms and their advancement deep into the crystal.

## II. SHOCK WAVES OF LONGITUDINAL DISPLACEMENTS

If the flat surface of an elastic medium at the initial moment of time receives a longitudinal impulse directed along the axis perpendicular to its surface, then quasi-one-dimensional longitudinal waves of atomic displacements arise in the medium, the evolution of which is described by the hyperbolic equation (Ref. 8)

$$u_{tt} = a^2 u_{xx} \qquad (1)$$

The function $u(x,t)$ represents at time $t$ the displacement of a point having an abscissa $x$ in the equilibrium position, $u_{tt}$ and $u_{xx}$ are the second derivatives of the displacements with respect to time and coordinate, $a=\sqrt{k/\rho}$ is the speed of sound in the rod, $k$ is Young's modulus, and $\rho$ is the density of the medium. The relative elongation of the sample at a point $x$ is equal to $u_x(x,t)$.
We will consider a long metal sample with length $l$, rigidly fixed at the origin of coordinates $\left(u(0,t)=0\right)$, which at time $t=0$ is in equilibrium $\left(u(x,t)=0,\ t\le 0\right)$. The free end of the sample $\left(u_x(l,t)=0\right)$ at the initial moment of time receives a longitudinal shock impulse $I$.

$$u_t(x,0)=-\frac{I}{\rho}\delta(x-x_0),\ \ 0<x_0<l,\ \ \ x_0\to l. \qquad (2)$$

To solve equation (1), we will use the method of separation of variables (Ref. 8), that put

$$u(x,t)=X(x)T(t). \qquad (3)$$

Substituting (3) into (1), taking into account (2), and, the above mentioned initial and boundary conditions, we obtain:

$$u(x,t)=-\frac{4I}{\pi a\rho}F(q,\eta), \qquad (4)$$

where

$$F(q,\eta)=\sum_{n=0}^{\infty}\frac{(-1)^n}{(2n+1)}\sin\left[\frac{(2n+1)\pi}{2}\eta\right]\sin\left[\frac{(2n+1)\pi}{2}q\right] \qquad (5)$$

Figure 1 shows the dependence of the reduced value of the displacement $u(x,t)/u_0$ on the dimensionless variables of time $q=\frac{a}{l}t$ and coordinates $\eta=\frac{x}{l}$, where $u_0=4I/(\pi a\rho)$.
From formulas (4), (5) and Fig. 1, it follows that after the impact, a deformation wave with a sharp front and constant amplitude is formed in the crystal, which propagates into the depth of the crystal without distortion. The speed of movement of the deformation wave front $u(x,t)$ is equal to $a$. The position of the shock front is determined by the relation $\eta_{Fr}=1-q$

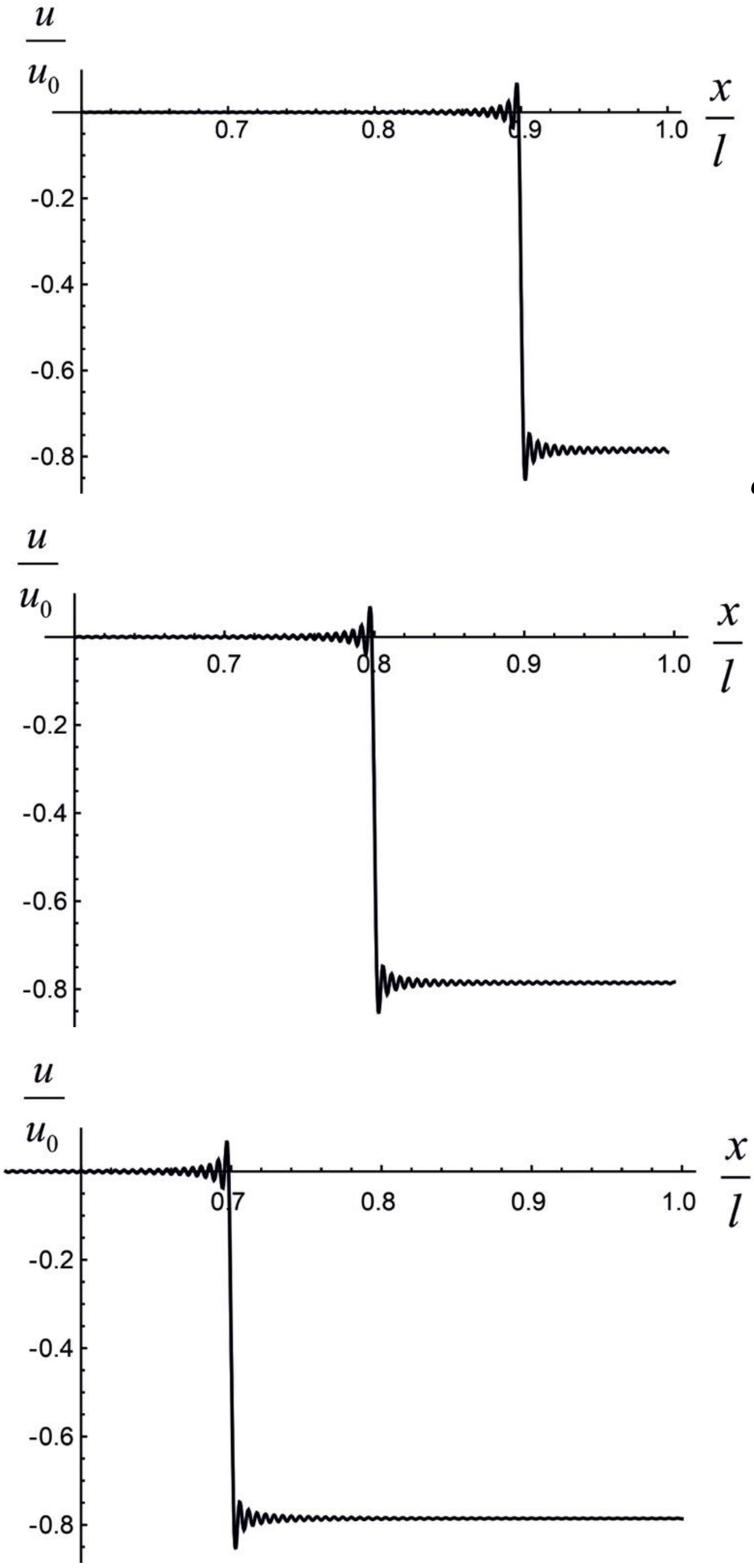

$q = 0.1$; $q = 0.2$; $q = 0.3$;

Fig. 1. Dependence of the reduced value of the displacement $u(x,t)/u_0$ on the dimensionless coordinate $\eta = \dfrac{x}{l}$ at different times.

## III. CHANGES IN THE MICROSTRUCTURE OF A CRYSTAL IN THE FIELD OF A SHOCK WAVE

Consider a metal rod with a crystal structure of a simple cubic lattice $\left(a_0 = b_0 = c_0\right)$. Let the lattice constant in the non-deformed state be $2a_0$ (see Fig.2).
After hitting the end surface of the bar $\left(x = l\right)$, the physical point of the bar, which at the initial moment occupies the position $x$, at any subsequent moment $t$ will be at the point with the coordinate $X\left(x,t\right) = x + u\left(x,t\right)$ (see e.g. Ref. 8).

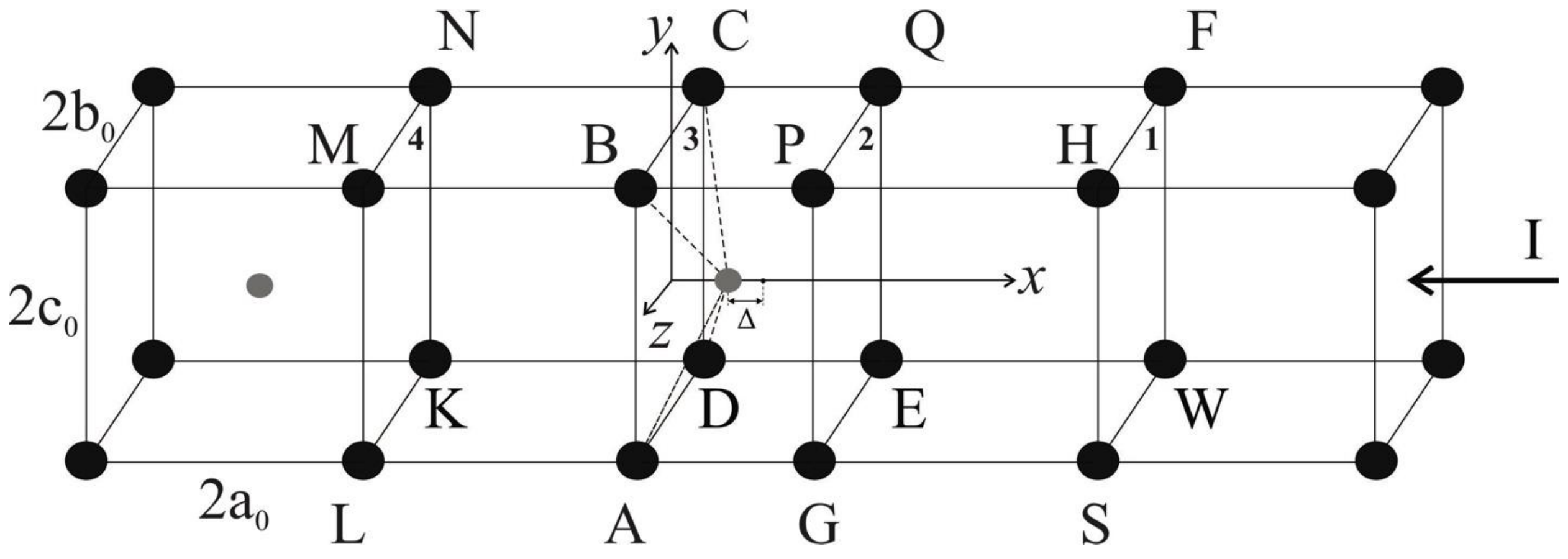

Fig. 2. Displacement of atoms of a cubic lattice during the passage of a deformation wave.

As a consequence, the lattice constant in the sample will be equal to

$$2a = 2a_0 + \left( u\left(x + 2a_0, t\right) - u\left(x,t\right) \right). \qquad (6)$$

In the region of uniform deformation of the sample $\left(x > G\right)$, $u\left(x+2a_0,t\right) \approx u\left(x,t\right)$, the lattice is not deformed $a = a_0$ (Fig. 2).
To the left of the front boundary $\left(x \le G\right)$ $u\left(x,t\right) = 0$, and the lattice constant at the wave front (AG layer in Fig. 2) is

$$2a_{AG} = 2a_0 + u\left(x_G, t\right). \qquad (7)$$

Since $u\left(x_G, t\right) < 0$, the interatomic distance in the layer of atoms, which is adjacent to the front of the deformation wave, decreases, i.e. the medium at the wavefront is compressed. Local dynamic compression of the medium (atomic layer AG in Fig. 2) will lead to a change in the potential relief and kinetic energy of the interstitial atom. This can be shown by direct calculation.
Let's choose the origin of coordinates in the center of the plane ABCD, in which we place the coordinate axes *yz*. The initial position of the interstitial atom will be determined by the coordinates $\left\{a_0 + u - \Delta, 0, 0\right\}$, where

$$X_{at} = a_0 + u - \Delta \qquad (8)$$

- the position of the interstitial atom on the $x$-axis, $\Delta$ - parameter that determines the current position of the atom. So, for example, in equilibrium when the atom is in the center of the cell $\left(X_{at,0} = a_0\right)$ $\Delta = 0$.
Let us assume that the interaction of an interstitial atom with matrix atoms is described by the Lenard-Jones potential

$$\mathrm{U}(\mathrm{R}_{s,j}) = 4\varepsilon\left[\left(\sigma / \mathrm{R}_{s,j}\right)^{12} - \left(\sigma / \mathrm{R}_{s,j}\right)^{6}\right], \quad (9)$$

where $\mathrm{R}_{s,j} = \left|\mathbf{R}_s - \mathbf{R}_j\right|$ is the distance between the interstitial atom $s$ and the j-th atom of the crystal.
The potential energy of an interstitial atom is determined by the energy of its interaction with the crystal atoms (9), and the transition of an atom under the influence of a deformation wave from the near-boundary crystal cell AG to the neighboring undeformed cell LA will be accompanied by a change in its potential energy.
We restrict ourselves to taking into account the interaction of the interstitial atom with the atoms of the matrix, which are located in 1, 2, 3, and 4 planes of the crystal lattice (Fig. 2). The distance between an interstitial atom and atoms located in the 1, 2, 3, and 4 planes of the crystal is

$$R(1) = a_0\sqrt{\left(3 + \Delta / a_0\right)^2 + 2}, \quad R(2) = a_0\sqrt{\left(1 + \Delta / a_0\right)^2 + 2}, \quad R(3) = a_0\sqrt{\left(1 + (u - \Delta) / a_0\right)^2 + 2}$$

$$R(4) = a_0\sqrt{\left(3 + (u - \Delta) / a_0\right)^2 + 2}. \quad (9)$$

The interaction energy of an interstitial atom with matrix atoms located in the 1st, 2nd, 3rd, and 4th planes of the crystal lattice is

$$\begin{aligned}\frac{\mathrm{U}(\chi,\gamma)}{16\varepsilon} = {} & s^{12}\left(\left[(3+\chi)^2 + 2\right]^{-6} + \left[(1+\chi)^2 + 2\right]^{-6} + \left[(1+\gamma-\chi)^2 + 2\right]^{-6}\right. \\ & \left. + \left[(3+\gamma-\chi)^2 + 2\right]^{-6}\right) \\ & - s^{6}\left(\left[(3+\chi)^2 + 2\right]^{-3} + \left[(1+\chi)^2 + 2\right]^{-3} + \left[(1+\gamma-\chi)^2 + 2\right]^{-3} + \left[(3+\gamma-\chi)^2 + 2\right]^{-3}\right),\end{aligned} \quad (10)$$

Where $\chi = \Delta / a_0$; $\gamma = u / a_0$; $s = \sigma / a_0$

In the equilibrium state before striking ( $I = 0$, $u = 0$ ), the distribution of the potential energy of an interstitial atom in neighboring cells LA and AG is symmetric with respect to the plane ABCD (Fig. 3).

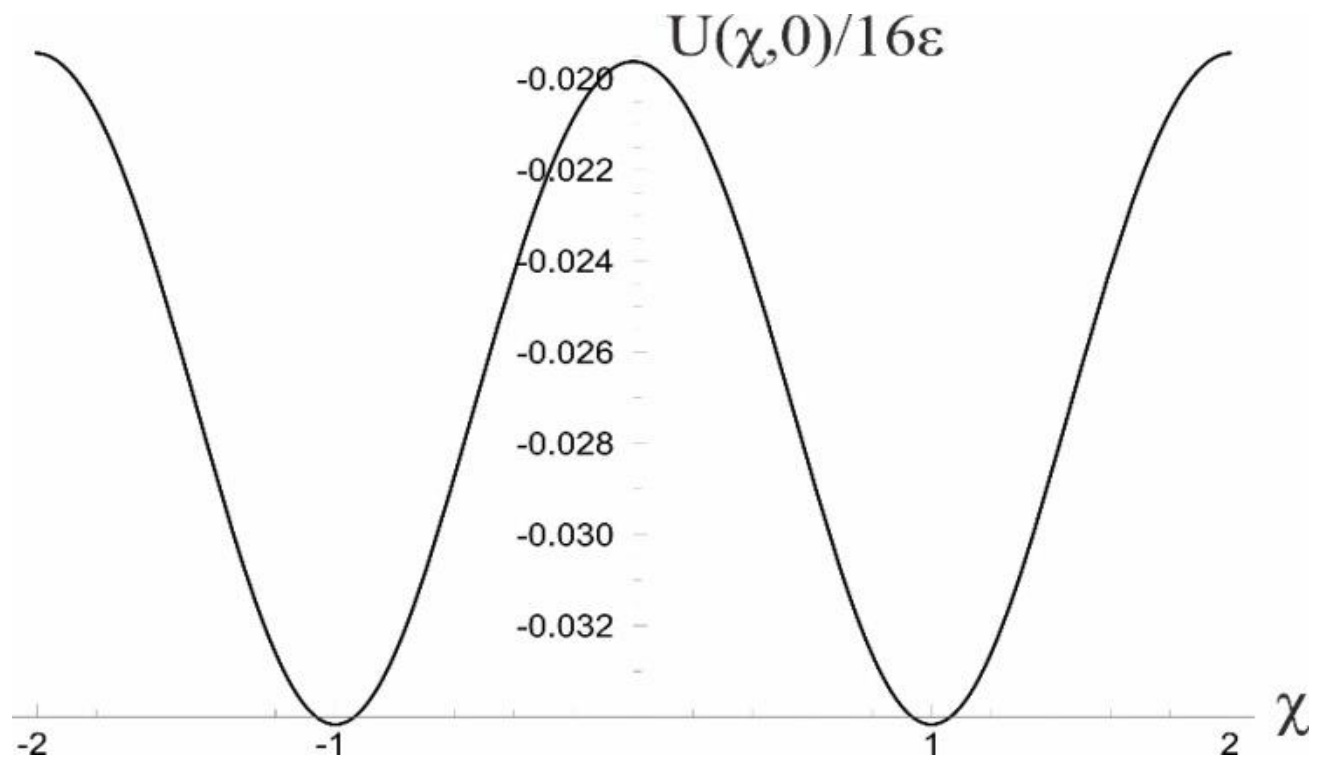

Fig. 3. Equilibrium distribution of the potential energy of an interstitial atom $I = 0,\ \ u = 0,\ s = 0.8$ .

After impact on the sample surface the potential barrier for the transition of atoms in the direction of impact decreases (Fig. 4)

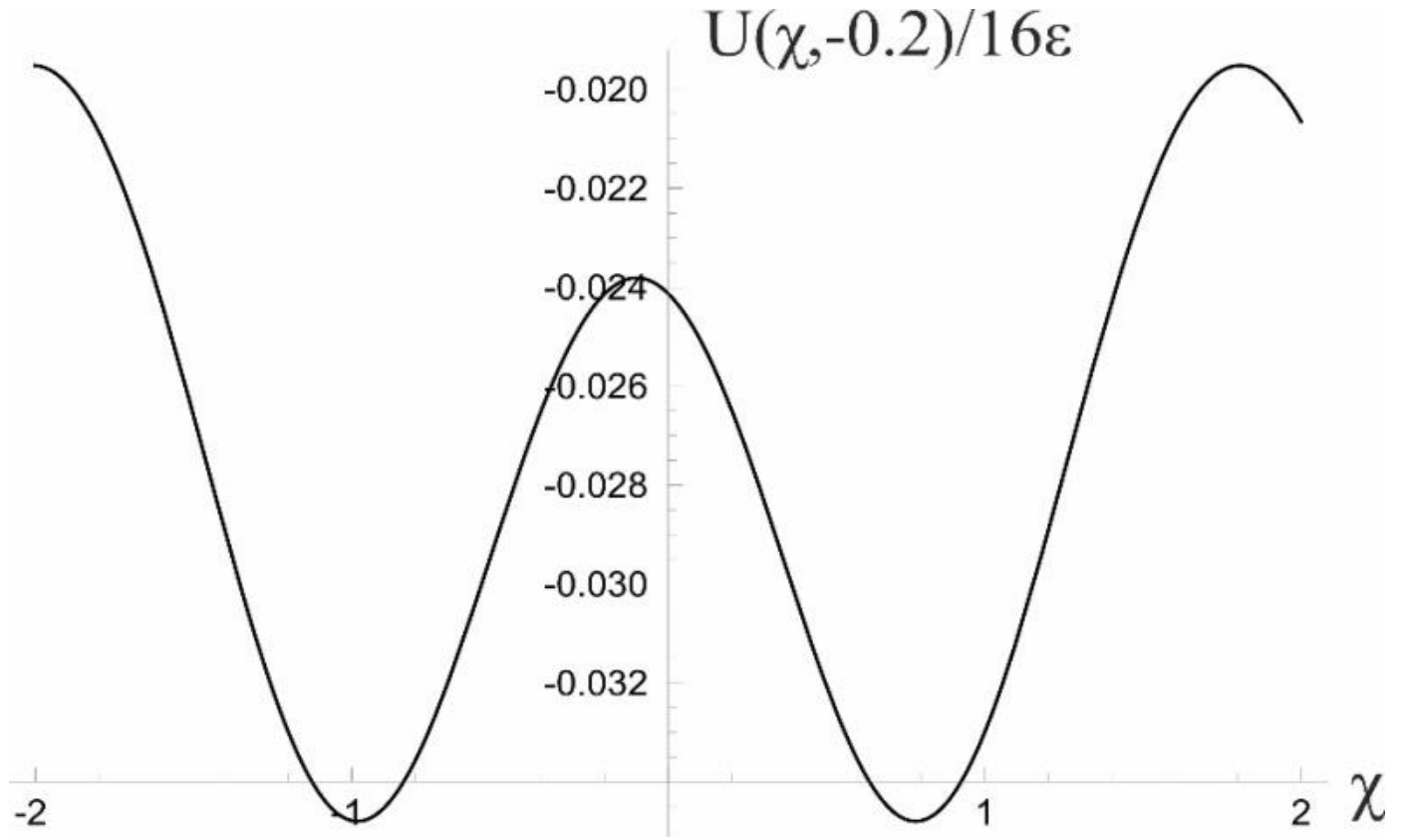

Fig. 4. At a small value of the deformation pulse $I$ ( $s = 0.8,\ u = -0.2$ ), the potential barrier to the transfer of interstitial atoms at the wave front decreases.

At a larger longitudinal momentum $I$ , the barrier separating the localized states of the interstitial atom practically disappears.

## IV. EMISSION FLUX OF INTERSTITIAL ATOMS IN THE FIELD OF A SHOCK WAVE

The impact deformation of the crystal surface is accompanied by the transfer of a part of the longitudinal momentum to interstitial atom. An increase in its kinetic energy makes it easier for the interstitial atom to overcome the potential barrier separating neighboring localized states. The speed acquired by an interstitial atom after the transfer of a shock pulse to the sample is equal to

$$V_{at} = \partial X_{at} / \partial t = \partial u(x,t) / \partial t . \tag{11}$$

Considering that $q = (a/l)t$,

$$\frac{\partial u(x,t)}{\partial t} = \frac{4I}{\pi \rho l} \frac{\partial F(q,\eta)}{\partial q} , \tag{12}$$

where $F(q,\eta)$ is determined by expression (5). To find the derivative $\partial F(q,\eta)/\partial q$ at the front of the deformation wave, $\eta_{Fr}=1-q_0$, $q_0=t_0\, a/l$ , the easiest way is to use numerical methods, assuming

$$\partial F(q,\eta)/\partial q\big|_{q\to q_0}=F(q_0+\Delta q)-F(q_0)/\Delta q\big|_{\Delta q\to 0} \quad (13)$$

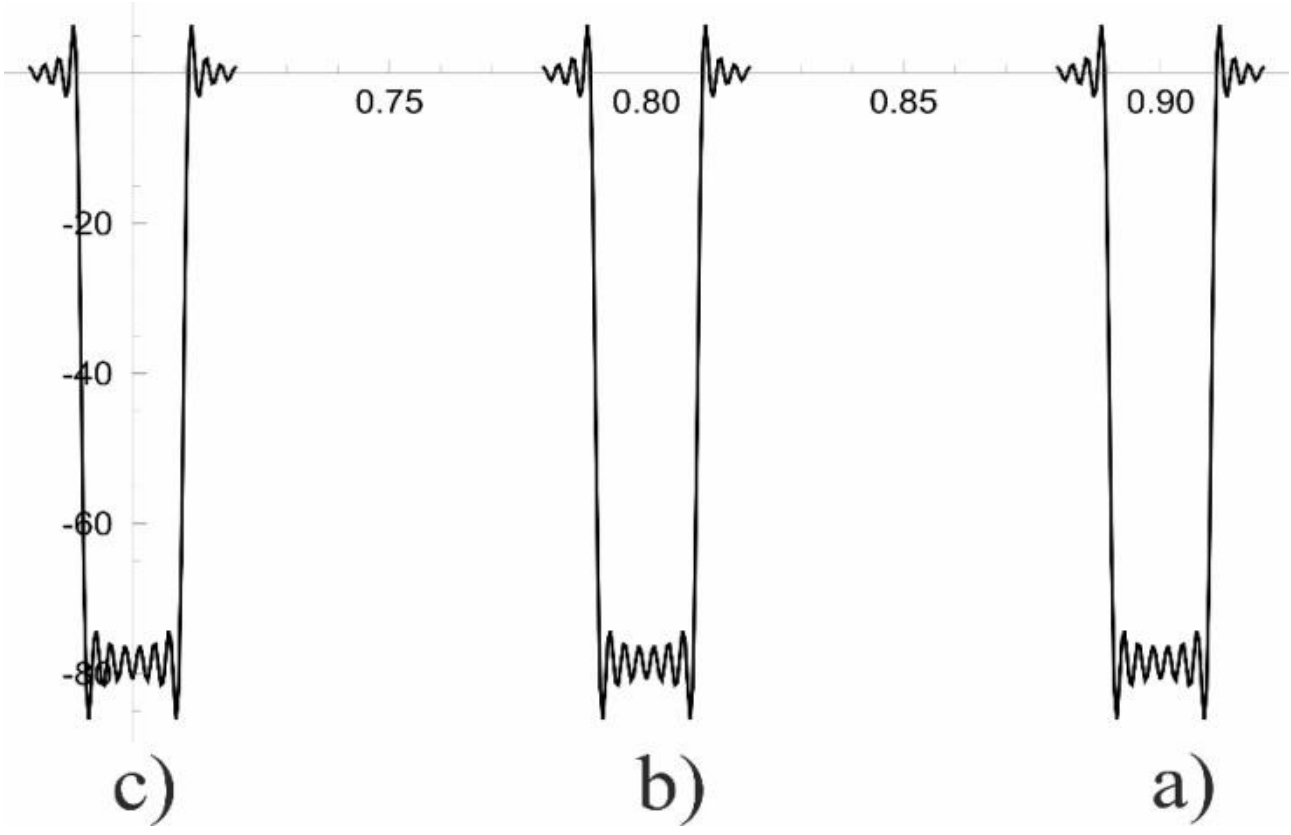

Fig. 5. The value of the derivative (13) at different positions of the deformation wave front: a) $\eta_{Fr}=0.1$; b) $\eta_{Fr}=0.2$; c) $\eta_{Fr}=0.3$.

As follows from (12) and Fig. 5, at the front of the deformation wave, the velocity of the longitudinal motion of the interstitial atom sharply increases and, consequently, its kinetic energy increases, the value of which can exceed the value of the activation barrier separating the minima of the potential energy of the interstitial atom in neighboring cells, for example, LA and AG. This will manifest itself in the activationless motion of the atom associated with the deformation wave. The additional kinetic energy that it receives, while the interstitial atom is equal to

$$E_\kappa=\frac{8mI^2}{\pi^2\rho^2 l^2}\left(\frac{\partial F}{\partial q}\right)^2. \quad (14)$$

Taking into account that $I=M_0V_0,\ \ \rho l=M$ , expression (14) can be rewritten as

$$E_k=\left(8/\pi^2\right)mV_0^2\left(M_0/M\right)^2\left(\partial F/\partial q\right)^2\Big|_{q\to q_0}, \quad (15)$$

As an example, we put

$$m=9.310^{-23}\,g,\ V_0=10^3\,cm/s, M_0/M=0.5,\ \left(\partial F/\partial q\right)_{q\to q_0}=-80,$$

than $E_k=0.4eV$, that is, the value of $E_k$ is comparable (or more) to the value of interstitial potential barriers. If the value $E_\kappa$ exceeds the value of the activation barrier that an interstitial atom needs to overcome in order to enter the neighboring cell, then the process of atom moving occurs without activation and emission of interstitial atoms into the bulk of the crystal occurs.

It will be emphasized that an increase in the longitudinal momentum leads not only to an increase in the kinetic energy of an interstitial atom, but also may significantly lowers the activation barrier for atomic transfer.

## V. CONCLUSIONS

As already noted, radiation action on the surface of a metal or semiconductor can lead to the escape of electrons or ions from the medium. This occurs when the energy of radiation, which is transferred to the charge, exceeds the work function of charges from the medium. A classic example of emission is the emission of electrons from metals and semiconductors, when, under the influence of an electromagnetic wave, part of the electrons passes into the conduction band (internal photoelectric effect) or leaves the medium (external photoelectric effect). A similar situation will also be observed in the case of the appearance of a deformation wave in a crystal containing interstitial atoms. Under impact action on the surface of the crystal, the atoms of the crystal are elastically displaced in layers, transferring the energy of elastic deformation deep into the crystal. The presence of an interstitial atom in the crystal lattice breaks the translational symmetry of the lattice and leads to the scattering (absorption) of the elastic wave energy by the interstitial atom. When the amount of absorbed energy exceeds the energy of localization of an interstitial atom in a crystal, the emission of an interstitial atom into the bulk of the crystal occurs.

## DATA AVAILABILITY

The data that supports the findings of this study are available within the article.